\documentclass[lettersize,journal]{IEEEtran}
\usepackage[switch]{lineno} % default option is 'left'

\usepackage[export]{adjustbox}

\title{Comparing Controller Synthesis Methods\\with Deadline-Miss Awareness}
\author{Anonymous author(s)}
\iftrue
\author{Melanie Gallant$^\star$, Marc Seidel$^\star$, Paolo Pazzaglia, Claudio Mandrioli,\\Christoph Mark, Kevin Schmidt, Frank Allg\"ower, Martina Maggio%
\thanks{$^\star$ The first two authors contributed equally to this work. 
M. Gallant (\emph{melanie.gallant@de.bosch.com}) is with Robert Bosch GmbH, Corporate Research, Germany and Saarland University, Germany.
M. Seidel (\emph{seidel@ist.uni-stuttgart.de}) is with University of Stuttgart, Institute for Systems Theory and Automatic Control, Germany.
P. Pazzaglia (\emph{paolo.pazzaglia@de.bosch.com}), C. Mark (\emph{christoph.mark@de.bosch.com}) and K. Schmidt (\emph{kevin.schmidt4@de.bosch.com}) are with Robert Bosch GmbH, Corporate Research,
Germany.
C. Mandrioli (\emph{claudio.mandrioli@uni.lu}) is with University of Luxembourg, Luxembourg.
F. Allg\"ower (\emph{allgower@ist.uni-stuttgart.de}) is with University of Stuttgart, Institute for Systems Theory and Automatic Control, Germany. 
M. Maggio (\emph{maggio@cs.uni-saarland.de}) is with Saarland University, Germany and Lund University, Sweden.\\  
This work was supported by:
the German Federal Ministry for Economic Affairs and Climate Action under Grant 13IPC021 (IPCEI-CIS), the European Union's Horizon Europe program under Grant Agreement No. 101148870 (ConTestCPS), the EU INTERREG North Sea project STORM\_SAFE of the European Regional Development Fund, and the ``Wallenberg AI, Autonomous Systems and Software Program (WASP) funded by the Knut and Alice Wallenberg Foundation'' via the NEST projects \emph{Intelligent Cloud Robotics for Real-Time Manipulation at Scale} and \emph{DYNACON: DYNamic Attack detection and mitigation for seCure AutONomy}.}%
\thanks{Manuscript received March 30, 2026; revised June 17, 2026; accepted July 17, 2026. 
}
}
\fi

\usepackage[scaled]{beramono}
\usepackage[T1]{fontenc}

\usepackage{cite}
\usepackage{booktabs}
\usepackage{amsmath}
\usepackage{amssymb}
\usepackage{mathtools}
\usepackage[table]{xcolor}
\usepackage{bbding} %for yes/no symbol
\usepackage{fontawesome5} %for dice symbol etc...
\usepackage{url}
\usepackage{multirow}
\usepackage{dblfloatfix}
\usepackage{cuted}
\usepackage[inline]{enumitem}
\usepackage{xspace}
\usepackage{threeparttable}
\usepackage{tikz}
\usepackage{subcaption}
\usepackage{newfloat}
\usepackage{makecell}
\usepackage{nicefrac}
\usepackage{tcolorbox}
\usepackage{tikz, pgfplots, pgfplotstable}
\usetikzlibrary{positioning}
\usetikzlibrary{decorations.pathreplacing}
\usepgfplotslibrary{fillbetween,groupplots}
\pgfplotsset{compat=1.10}

\usepackage{multicol}
\usepackage[acronym,shortcuts]{glossaries}
\makeglossaries
\usepackage{hyperref}

\usepackage{titlesec}
\titlespacing*{\section}{0pt}{0.9ex}{0.9ex}
\titlespacing*{\subsection}{0pt}{0.9ex}{0.9ex}

\DeclareFloatingEnvironment[
    fileext=lob, % List of Boxes file extension
    name=Box,    % The name to be printed in the caption
    placement=htbp  % Default placement options (here, top, bottom, page)
]{myfigure}

\newcommand{\PlantNonlinear}{\mathcal{P}}
\newcommand{\PlantLinearContinuousTime}{\PlantNonlinear_{\mathrm{LTI\,ct}}}
\newcommand{\PlantLinearDiscreteTime}{\PlantNonlinear_{\mathrm{LTI\,dt}}}

\newcommand{\Acont}{A}
\newcommand{\Bcont}{B}
\newcommand{\Ccont}{C}
\newcommand{\Dcont}{D}
\newcommand{\Adisc}{\Phi}
\newcommand{\Bdisc}{\Gamma}
\newcommand{\Cdisc}{\Ccont}
\newcommand{\Ddisc}{\Dcont}

\newcommand{\paperNameStreamAdapt}{\texttt{StrAdapt}}
\newcommand{\paperNameGaid}{\texttt{AccControl}}
\newcommand{\paperNameMS}{\texttt{GSS}}
\newcommand{\paperNameMPC}{\texttt{TreeMPC}}
\newcommand{\paperNameDMAC}{\texttt{DMAC}}
\newcommand{\paperNameMJLS}{\texttt{DR-MJLS}}
\newcommand{\paperNameVreman}{\texttt{DMAI}}
\newcommand{\paperNameHorssen}{\texttt{GraphLQR}}

\newcommand{\bigO}[1]{\mathcal{O}(#1)}

\newcommand{\dimState}{\ensuremath{n_\mathrm{x}}}
\newcommand{\dimInput}{\ensuremath{n_\mathrm{u}}}
\newcommand{\dimOutput}{\ensuremath{n_\mathrm{y}}}
\newcommand{\dimCtrState}{\ensuremath{n_\mathrm{z}}}
\newcommand{\dimExtendedState}{\ensuremath{n_\mathrm{xu}}}
\newcommand{\extendedState}{\tilde{x}}
\newcommand{\Nodes}{\mathcal{V}}
\newcommand{\Edges}{\mathcal{E}}
\newcommand{\numNodes}{\ensuremath{n_{\mathcal{V}}}}
\newcommand{\numEdges}{\ensuremath{n_{\mathcal{E}}}}
\newcommand{\wcrt}{R_\mathrm{max}}
\newcommand{\step}{k}
\newcommand{\policy}{\kappa}
\newcommand{\period}{T}
\newcommand{\periodSensor}{T_\mathrm{s}}
\newcommand{\numControllers}{c}

\newcommand{\Kill}{\textsc{Kill}}
\newcommand{\Skip}{\textsc{SkipNext}}
\newcommand{\Queue}{\textsc{Queue}}
\newcommand{\Zero}{\textsc{Zero}}
\newcommand{\Hold}{\textsc{Hold}}

\newcommand{\stratov}{\ensuremath{s_\mathrm{o}}}
\newcommand{\stratovset}{\ensuremath{\mathcal{S}_\mathrm{o}}}
\newcommand{\stratac}{\ensuremath{s_\mathrm{a}}}
\newcommand{\stratacset}{\ensuremath{\mathcal{S}_\mathrm{a}}}

\newcommand{\hit}{\textsc{h}}
\newcommand{\miss}{\textsc{m}}
\newcommand{\recovery}{\textsc{r}}

\newcommand{\nhit}{h}
\newcommand{\nmiss}{m}
\newcommand{\window}{w}

\newcommand{\anyHit}[2]{\ensuremath{\textbf{\texttt{AnyHit}}( {#1, #2})}}
\newcommand{\anyMiss}[2]{\ensuremath{\textbf{\texttt{AnyMiss}}( #1, #2)}}
\newcommand{\rowHit}[2]{\ensuremath{\textbf{\texttt{RowHit}}( #1, #2)}}
\newcommand{\rowMiss}[1]{\ensuremath{\textbf{\texttt{RowMiss}}( #1)}}

\newcommand{\yes}{\CheckmarkBold}
\newcommand{\no}{\XSolidBrush}
\newcommand{\iconStochastic}{\faDice}
\newcommand{\iconDeterministic}{\faCogs}
\newcommand{\matvecmultOrig}[2]{\tikz{\draw[thick,black,fill=black!30!white] (0,0) rectangle (0.3cm,0.3cm);}$_{\,#1}$\,$^{\dotp}$\,\tikz{\draw[black,thick,fill=black!30!white] (0,0) rectangle (0.1cm,0.3cm);}$_{\,#2}$}
\newcommand{\matvecmult}[3]{\matvecmultOrig{#1 \times #2}{#3}}

\newacronym{mpc}{MPC}{Model Predictive Control}
\newacronym{mjls}{MJLS}{Markov Jump Linear System}
\newacronym{lmi}{LMI}{Linear Matrix Inequality}
\newacronym{ocp}{OCP}{Optimal Control Problem}
\newacronym{socp}{SOCP}{Second-Order Cone Program}
\newacronym{sdp}{SDP}{Semi-Definite Program}
\newacronym{wh}{WH}{Weakly-Hard}
\newacronym{lti}{LTI}{Linear Time-Invariant}
\newacronym{lqr}{LQR}{Linear-Quadratic Regulator}
\newacronym{jsr}{JSR}{Joint Spectral Radius}
\newacronym{wcrt}{WCRT}{Worst-Case Response Time}
\newacronym{zoh}{ZOH}{Zero-Order Hold}
\newacronym{lqg}{LQG}{Linear Quadratic Gaussian}
\newacronym{let}{LET}{Logical Execution Time}
\newacronym{ncs}{NCS}{Networked Control Systems}
\newacronym{sre}{SRE}{Stochastic Riccati Equation}
\newacronym{are}{ARE}{Algebraic Riccati Equation}
\newacronym{pmsm}{PMSM}{Permanent Magnet Synchronous Motor}

\tikzstyle{block} = [draw, rectangle]
\tikzstyle{matrix} = [block, minimum height=3em, minimum width=3em]
\tikzstyle{vector} = [block, rotate=90, minimum height=1.333em, minimum width=3em]
\newcommand{\dotp}{
    \mathop{
        \mathchoice{\vcenter{\hbox{\huge$\cdot$}}}
                   {\vcenter{\hbox{\LARGE$\cdot$}}}
                   {\vcenter{\hbox{\large$\cdot$}}}
                   {\vcenter{\hbox{\small$\cdot$}}}
    }
}

\renewcommand{\paragraph}[1]{\noindent{\bf#1}:}

\definecolor{firstcol}{HTML}{EE7733}
\definecolor{secondcol}{HTML}{0077BB}
\definecolor{thirdcol}{HTML}{99ddf6}
\definecolor{fourthcol}{HTML}{EE3377}
\definecolor{fifthcol}{HTML}{CC3311}
\definecolor{sixthcol}{HTML}{009988}
\definecolor{seventhcol}{HTML}{BBBBBB}
\definecolor{offwhite}{HTML}{FAEBD7}

\colorlet{colfirststate}{firstcol}
\colorlet{colsecondstate}{fifthcol}
\colorlet{colthirdstate}{seventhcol}
\colorlet{colfirstinput}{secondcol}
\colorlet{colsecondinput}{thirdcol}

\newcommand\copyrighttext{%
  \footnotesize \textcopyright 2026 IEEE. Personal use of this material is permitted. Permission from IEEE must be obtained for all other uses, including reprinting/republishing this material for advertising or promotional purposes, collecting new collected works for resale or redistribution to servers or lists, or reuse of any copyrighted component of this work in other works.}
\newcommand\copyrightnotice{%
\begin{tikzpicture}[remember picture,overlay]
\node[anchor=south,yshift=5pt] at (current page.south) 
  {\fbox{\parbox{\dimexpr\textwidth-\fboxsep-\fboxrule\relax}{\copyrighttext}}};
\end{tikzpicture}%
}

\usetikzlibrary{external}

\begin{document}

\maketitle
\copyrightnotice
\begin{abstract}
This paper presents a comparative study of real-time controller design methods for dynamical control systems subject to deadline overruns. 
Numerous techniques have been proposed to handle missed deadlines, i.e., including models or information of possible deadline misses directly in the control design.
These approaches substantially differ in their assumptions, supported execution semantics, required runtime information, and provided guarantees, making direct comparison challenging. 
We address this gap with a qualitative assessment of representative deadline-miss-aware control methods.
We analyze the design space of such controllers and identify the critical assumptions on the real-time constraints, the control design criteria and implementation aspects.
We expose how these different assumptions lead to the different trade-offs in applicability, and control performance and guarantees.
We contribute a comparison tool enabling researchers to  systematically benchmark new techniques against the state of the art. 
The study is supported by simulations on two case studies: a Furuta pendulum and an electric motor to illustrate the practical behavior of the different approaches under missed-deadline conditions. 

\end{abstract}

\section{Introduction}
\label{sec:intro}

Control systems shape our lives in many different domains: from transportation to home appliances.
These systems typically have to operate in real-time, providing answers and interacting with physical environments.
Industrial studies report that many of these systems experience violations of the assumptions on the timing of the software execution, i.e., occasional deadline misses~\cite{akesson2022comprehensive} that sometimes could be tolerated, at least if sufficiently rare~\cite{cervin2005overrun}.
They may result from transient overloads, execution-time variability, or communication delays~\cite{maggio2020control}, and, as they may deteriorate performance or affect safety, should be a fundamental design concern.

Disregarding this consideration, most control applications are still developed under the assumption that functional and timing aspects can be addressed separately. 
This perspective underlies abstractions such as the classic hard-deadline contract, in which missing a deadline is considered a system failure. Consequently, functional correctness is reduced to ensuring that all tasks remain schedulable.
This abstraction has enabled clean design processes, but in practice seems to be overly conservative. 
Consequences are either over-provisioning computational resources to improve response times, or selecting controller periods large enough to guarantee deadline satisfaction in all conditions. 
The former carries an obvious implementation cost, whereas the latter can significantly degrade control quality.
Larger sampling periods lead to worse disturbance rejection, worse performance, and reduced stability margins~\cite{cervin2003does}.
Controllers with shorter periods, that tolerate occasional deadline misses, achieve a better trade-off between implementation cost and control quality~\cite{pazzaglia2019dmac}.

From a control perspective, timing effects constitute a major source of uncertainty, alongside plant disturbances and modeling errors. 
In software-based control systems executed on shared real-time platforms, these effects arise naturally from inherent execution-time variability, interference from co-running tasks, access to data in memories and scheduling decisions. 
The resulting delays are typically time-varying and may exceed one sampling interval~\cite{aastrom2013computer}.
Additional design choices are then required to handle overlapping job instances, as well as delayed or missed control updates~\cite{cervin2003does}. 

These issues have motivated a broad body of work on controller synthesis and strategies that explicitly account for deadline overruns, by designing a specific control law that either is more robust to this type of disturbances or changes depending on the history of past deadline outcomes.
However, the available techniques differ substantially in their assumptions, objectives, and intended operating conditions. 
They may rely on different models of timing uncertainty, actuator and scheduling semantics, and levels of runtime information. 
As a result, it is often difficult to understand how methods relate to one another, in which scenarios they are applicable, and what trade-offs impose in terms of guarantees, complexity, and practical deployability. 
This lack of a common comparison basis makes it harder both for practitioners to choose an appropriate method and for researchers to position new contributions against prior work.

In this paper, we address this gap and provide a qualitative comparison of deadline-miss-aware controller design techniques.
We identified eight methods~\cite{gaid2008dropin, vanhorssen2016performance, pazzaglia2019dmac, pazzaglia2021adaptive, vreman2022adaptive, gallant2024structure, gallant2025soft, seidel2026framework} that, to the best of our knowledge, represent the current snapshot of state of the art that explicitly account for deadline misses in the controller design.
Each approach is based on different assumptions on 
\begin{enumerate*}[label=(\roman*)]
    \item the real-time system,
    \item control design criteria, and
    \item implementation choices.
\end{enumerate*}
Such variety leads to different trade-offs in applicability, and control performance and guarantees.
Our goal is to  identify how the different assumptions of the design approaches generate these trade-offs, and thus to their design space.
While there is not one best design approach overall, with our work we offer valuable directions to researchers and practitioners, enabling them to choose the right approach for their given assumptions and required guarantees, as well as identifying gaps for future research directions.
To complement the qualitative analysis, we propose a comparison tool that allows to consistently integrate and evaluate new techniques against the literature.
We illustrate its usage by providing simulations based on two systems subject to deadline overrun conditions: a Furuta pendulum, a widely used benchmark for stress-testing control designs~\cite{Gafvert1998modelling}, and an electric motor as an industrially-relevant use-case~\cite{maggio2020control}.
We show simulation results of all deadline-miss-aware controllers, as well as specific examples that underline the insights from the qualitative analysis.
We evaluate the control performance of different designs under varying deadline-miss probabilities and under partial state measurements.
Our results expose the approaches' diverse trade-offs and targeted applications, showing the design decisions that define the targeted context, providing guidance for future research.

\section{Background and basic notions}
\label{sec:basics}

This section introduces concepts and notation that we will use in the rest of the paper.
Specifically, Sec.~\ref{sec:basics:plant} introduces the system model, Sec.~\ref{sec:basics:misses} describes how the system may react to deadline overruns, Sec.~\ref{sec:basics:missmodel} provides an overview of how to model patterns of deadline overruns, and Sec.~\ref{sec:basics:controller} gives some background on controller types.

%%%%%%%%%%%%%%%%%%%%%%%%%%%%%%%%%%%%%%%%%%%%%%%%%%%%%%%%%%%%%%%%%%%%%%%%%%%%%%%%%%%%%%%%%%%%%%%%%%%%

\subsection{System model}
\label{sec:basics:plant}

We define the physical plant to be controlled.
We make use of both a continuous-time nonlinear model

\begin{equation}
\label{eq:plant:nonlinear}
\PlantNonlinear \coloneq \left\{\,\, \begin{aligned}
\dot{x}(t) &= f\left( x(t), u(t) \right) \\
      y(t) &= g\left( x(t), u(t) \right) \\
\end{aligned} \right. ,
\end{equation}
where $f(\cdot, \cdot)$ and $g(\cdot, \cdot)$ are given functions that describe the plant dynamics, and its \gls{lti} counterpart, i.e., the model that is obtained linearizing Eq.~\eqref{eq:plant:nonlinear} around an operating point~\cite{Mandrioli:2024a},
\begin{equation}
\label{eq:sys-ct}
\PlantLinearContinuousTime \coloneq \left\{\,\, \begin{aligned}
\dot{x}(t) &= \Acont\,x(t) + \Bcont\,u(t) \\
      y(t) &= \Ccont\,x(t) + \Dcont\,u(t) \\
\end{aligned} \right. .
\end{equation}
In both equations, $x(t) \in \mathbb{R}^{\dimState}$ is the system state at time $t$, $u(t) \in \mathbb{R}^{\dimInput}$ is the system input, and $y(t) \in \mathbb{R}^{\dimOutput}$ is the system output, with $\dimState$, $\dimInput$, and $\dimOutput$ respective vector dimensions.
The matrices $\Acont, \Bcont, \Ccont$, and $\Dcont$ have appropriate size and describe the evolution of the linearized system in continuous-time.
We assume $t \geq 0$, $x(0) \coloneq x_0$ and $u(0) \coloneq u_0$.

The system input is calculated by a controller that acts periodically every $\period\,$seconds and applied using a zero-order-hold policy (i.e., the control signal is kept constant during the control period).
The controller is designed to guarantee stability (i.e., convergence of the system state) and maximize performance.
The performance is usually expressed as a {\em cost function}, which captures objectives such as convergence speed and control effort.
The controller is implemented as a real-time task, running on a platform that is possibly shared with other applications. An execution of the controller task, called \emph{job}, should be completed within its corresponding real-time \emph{deadline}. When this happens, we call it a \emph{deadline hit}, while an execution not completing within the deadline is referred to as a \emph{deadline overrun} or \emph{deadline miss}.

Under ideal conditions (i.e., with no deadline overruns), the controller input of the $\step$-th iteration of the controller task is sampled periodically at $\step \period, \step \in \mathbb{N}_0$.
Denoting with $x[\step]$ the value of the plant state at time $\step \period$, i.e., $x[\step] \coloneq x(\step \period)$, we obtain the discrete-time counterpart of Eq.~\eqref{eq:sys-ct} as
\begin{equation}
\label{eq:sys-dt}
\PlantLinearDiscreteTime \coloneq \left\{\,\, \begin{aligned}
x[\step + 1] &= \Adisc\,x[\step] + \Bdisc\,u[\step] \\
    y[\step] &= \Cdisc\,x[\step] + \Ddisc\,u[\step] \\
\end{aligned} \right. ,
\end{equation}
where the matrices $\Adisc$ and $\Bdisc$ are defined as~\cite{aastrom2013computer}
\begin{equation}
\label{eq:discretization}
\Adisc \equiv \Adisc(\period) = e^{\Acont\period}, \quad \Bdisc \equiv \Bdisc(\period) = \int\nolimits_{0}^{\period} e^{\Acont\nu} \mathrm{d}\nu \, \Bcont.
\end{equation}

We compare controllers compatible with the \gls{let} paradigm~\cite{kirsch2012let}, i.e., where the control task reads its input $x[\step]$ at the start of each period and applies the calculated control signal (i.e., the plant input) at the end of each period.
In control terms, this corresponds to the design of a one-step delay controller.
The system that incorporates a one-step delay controller is usually formulated using the augmented state $\extendedState[\step] = [x[\step]^\top u[\step]^\top]^\top$ with size $\dimExtendedState=\dimState+\dimInput$.

In this paper, we compare controllers that 
implement or adapt a \emph{control policies} $\policy$.
The information used by the control policy creates a distinction between different classes of controllers, which we discuss in Sec.~\ref{sec:basics:controller}.
In general, the result of the execution of the control policy~$\policy[\step]$ at time step $\step$ is the control signal~$u[\step+1]$. 
However, deadline overruns may prevent the policy from providing the next value of $u$, as described in the following.

%%%%%%%%%%%%%%%%%%%%%%%%%%%%%%%%%%%%%%%%%%%%%%%%%%%%%%%%%%%%%%%%%%%%%%%%%%%%%%%%%%%%%%%%%%%%%%%%%%%%

\subsection{Modeling the effect of deadline overruns}
\label{sec:basics:misses}

Under ideal conditions, the execution of $\policy[\step]$ reads its input, e.g., $x[\step]$, at $\step\period$ and produces its output $u[\step+1]$ within time $(\step+1)\period$.
From the real-time perspective, this is guaranteed when the control period $\period$ is selected to ensure that $\period \geq \wcrt$ where $\wcrt$ is the control task's \gls{wcrt}.
However, selecting $\period \geq \wcrt$ often leads to controllers that are too slow in reacting to changes in the environment~\cite{cervin2003does}.
In fact, the selection of $\period$ is constrained by physical considerations, and the designer has to accept that the control task may occasionally overrun~\cite{akesson2022comprehensive, maggio2020control}, causing a deadline miss.
When a job misses its deadline, the operating and runtime system can deal with its remaining execution in different ways, i.e., they can use different \textit{overrun strategies}~\cite{cervin2005overrun}.
The most common are:

\begin{enumerate}[label=(\roman*)]
\item \Kill{}: The job that overruns the deadline is aborted at its deadline, all the calculations performed up to the deadline are discarded, and the internal state of the controller is reset to the one it had at its activation. 
The following job is activated normally.
\item \Skip{}: The job that overruns the deadline is executed until its computation is completed. 
Any further job that, under nominal conditions, would have been activated while the computation is ongoing is discarded. 
A new job is activated only at the next release instant after the completion of the overrunning job.
\item \Queue{($i$)}: The job that overruns the deadline is executed until its computation is completed. 
Any job that, under nominal conditions, would have been activated while the computation is ongoing is placed in a queue and started (in FIFO order) immediately after completion of the overrunning job.
The queue only keeps the most recent $i$ jobs.
\Queue{(1)} is the most common case~\cite{pazzaglia2019dmac, maggio2020control} (will also be referred hereafter as \Queue{} for compactness).
\end{enumerate}

In this paper, we consider the three standard overrun strategies, formally grouped in the set $\stratovset \coloneq\{$\Kill{}, \Skip{}, \Queue{}$\}$, and we
define the function $\mu[\step]: \mathbb{N}_0 \to \left\{ \hit, \recovery, \miss \right\}$ which encodes the \emph{outcome} of the job execution of the control task at each time step $k\in\mathbb{N}_0$.
Having $\mu[\step]=\hit$ (``hit'') implies that the job either activated or enqueued at time $\step\period$ meets its deadline, which is set at time $(\step+1)\period$.
Having $\mu[\step]=\recovery$ (``recovery'') implies that a job activated at time $(\step-d)\period < \step\period$, for some $d>0$, terminates its computation in the interval $(\step\period,(\step+1)\period]$.
For periods where no new control signal is generated, i.e., with neither a timely nor a late termination of a control job, the outcome of the period is denoted with $\mu[\step]=\miss$, implying a deadline miss.
Under \Skip{} and \Queue{(1)}, all three outcomes are possible, while \Kill{} allows only \hit{} and \miss{}, since any job missing its deadline is discarded and no late termination may occur.

When a deadline miss of a control task  occurs, the system does not have a new control signal $u[\step+1]$ to actuate, and there are various alternatives for how to decide what value is assigned to the actuators.
We refer to this decision as \emph{actuator strategy}.
Common choices in the literature~\cite{cervin2005overrun, schenato2009zero} are \Hold{} and \Zero{}.
The former implies that the input applied to the system remains constant, i.e., $u[\step+1] \coloneq u[\step]$.
The latter implies that $u[\step+1] \coloneq 0_{\dimInput}$.
These enforcements are possibly repeated across multiple periods until a new control signal is calculated.
Simulations and experimental results in literature \cite{schenato2009zero}, \cite{vreman2021stability} show that the actuator strategy may have a significant impact on the control performance and the best choice depends on the plant dynamics.
We refer hereafter to the set of actuator strategies as $\stratacset \coloneq \{ $\Hold{}, \Zero{}$ \}$.

The value of $u[\step+1]$, applied starting at time $(\step+1)T$, depends on the freshest completed job and on the chosen actuator strategy $\stratac \in \stratacset$.
Additionally, based on the chosen overrun strategy $\stratov \in \stratovset$, such job might have been activated more than one period before.
The coupling of the actuator input and the control signal can be formally defined as 
\begin{equation*}
    u[\step+1] = \begin{cases}
        \policy[\step] & \mu[\step] = \hit \\
        u[\step] & \mu[\step] = \miss \land \stratac = \Hold{}\\
        0_{\dimInput} & \mu[\step] = \miss \land \stratac = \Zero{}\\
        \policy[\step-d] &  (\mu[i])_{i\in\{\step-d-1,\dots,k\}} \in \mathcal{R}_{\stratov}(d),
    \end{cases}
\end{equation*}
where, for $\stratov =$ \Skip{},
\begin{align*}
    \mathcal{R}_{\Skip{}}(d) = \{ &(\mu_0,\dots,\mu_{d+1}) \mid
    \mu_0 \in \{\recovery, \hit \} \\ 
    & \land \, \mu_i = \miss, \,\,\forall i \in \{ 1,\dots, d\}  \, \land \, \mu_{d+1} = \recovery \},
\end{align*}
and, for $\stratov =$ \Queue{},
\begin{align*}
    \mathcal{R}_{\Queue{}}(d) = \{ &(\mu_0,\dots,\mu_{d+1}) \mid \mu_{d+1} = \recovery \, \land \\
     & \quad [(\mu_0 = \hit \land \mu_i = \miss\,\, \forall i \in \{1,\dots, d\} ) \\
    & \quad \; \vee (\mu_1 = \recovery \land \mu_i = \miss\,\, \forall i \in \{2,\dots, d \})] \}.
    \end{align*}%
Based on the selected overrun and actuator strategy, the possible sequences of missed and hit deadlines that the control task may experience impact the closed-loop system differently, possibly affecting also the feasibility of designing a control policy that guarantees closed-loop properties such as stability.
%%%%%%%%%%%%%%%%%%%%%%%%%%%%%%%%%%%%%%%%%%%%%%%%%%%%%%%%%%%%%%%%%%%%%%%%%%%%%%%%%%%%%%%%%%%%%%%%%%%%
%
\subsection{Modeling the patterns of deadline overruns}
\label{sec:basics:missmodel}

Controllers that are aware of deadline overruns and actively react to these events often use knowledge about their occurrence, usually encoded as patterns of deadline hits and overruns.
The models of such patterns can be classified in two categories: \emph{probabilistic} and \emph{deterministic}.

Probabilistic schedulability research provides methods for quantifying or bounding deadline-miss rates and deadline-miss probabilities, potentially based on \gls{wcrt} analysis~\cite{davis2019schedulability}.
There are many factors that are causing overload conditions, such as variable execution times, interference from other tasks, cache misses, etc.---and their interplay is not obvious.
This makes the exact computation of a probabilistic model nontrivial~\cite{Davis2019b, Markovic2024, Antunes2021}.
The related works considered in this paper use tools such as scenario theory and sampling-based statistical analysis to provide a probabilistic model of deadline hit/miss patterns.
Frequently, the observed probabilistic patterns are modeled using Markov chains and their corresponding Markov graphs~\cite{vreman2023stochastic, gallant2025soft} and represented as probability-based transition matrices.
When employed in a \gls{mjls} framework, these models facilitate the analysis and control of systems with stochastically switching dynamics.

Deterministic models describe properties that the patterns of deadline outcomes should satisfy.
The commonly used deterministic models are the $(\nhit,\window)$-firm\footnote{In the literature the $(\nhit,\window)$-firm model is usually referred to as $(m,k)$-firm model. We renamed the variables for consistency as we use $m$ to indicate deadline misses, and $k$ for the discrete time step.} model~\cite{Hamdaoui1995} and \gls{wh} constraints~\cite{bernat2001weakly}, as synthesized below:
\begin{enumerate}[label=(\roman*)]
\item \anyHit{\nhit}{\window} or $(\nhit,\window)$-firm: 
in any $\window$ consecutive outcomes there are at least $\nhit$ successful ones (\hit{} or \recovery{}). 
\item \anyMiss{\nmiss}{\window}: 
%the model prescribes that 
in any $\window$ consecutive outcomes there are at most $\nmiss$ deadline overruns (\miss{}). This model is related to the $(\nhit,\window)$-firm model with $\anyMiss{\nmiss}{\window} \equiv \anyHit{\window-\nmiss}{\window}$.
\item \rowHit{\nhit}{\window}: in any $\window$ consecutive outcomes there are at least $\nhit$ \emph{consecutive} successful ones (\hit{} or \recovery{}).
\item \rowMiss{\nmiss}:  
in any window there are at most $\nmiss$ \emph{consecutive} deadline overruns (\miss{}).
Originally defined with a window, this model is actually independent of the window size and $\rowMiss{\nmiss} \equiv \anyMiss{\nmiss}{\nmiss+1}$.
\end{enumerate}
Each of the deterministic models above corresponds to a \emph{regular language}.
This can be proved formally by first encoding every job outcome over the binary alphabet $\Sigma = \{0,1\}$, where $0$ is a deadline miss (\miss{}) and $1$ a successful outcome (\hit{}, \recovery{}). 
A sequence of task executions is a word $w \in \Sigma^*$, and a deterministic model $\lambda$ defines the language $\mathcal{L}(\lambda) = \{w \in \Sigma^* \mid w \models \lambda \}$ of all finite outcome sequences that satisfy it. 
Constraints (i)--(iv) are \emph{regular} languages because all of them are \emph{sliding-window} properties of bounded memory: whether a word $w$ continues to satisfy $\lambda$ after appending a new outcome $c \in \Sigma$ depends only on a bounded-length suffix of $w$.
For \anyHit{\nhit}{\window} and equivalently \anyMiss{\nmiss}{\window}, the acceptance of the next outcome is determined entirely by the last $\window-1$ outcomes, since the constraint is re-evaluated on every window of $\window$ consecutive jobs. 
For \rowHit{\nhit}{\window} the same bound applies, while for \rowMiss{\nmiss} it suffices to remember the number of trailing consecutive overruns, which is capped at $\nmiss$ and hence also finite. In every case, the set of suffixes that the automaton must distinguish is finite.
By the Myhill-Nerode theorem~\cite{hopcroft06}, $\mathcal{L}(\lambda)$ is thus regular and recognized by a finite-state machine whose states are exactly the relevant bounded suffix of the outcome history and whose transitions are labeled by the feasible next outcomes in $\Sigma$.

Regular languages can be represented using directed graphs (see Box~\ref{box:graph}).
Graph edges either correspond to a single outcome~\cite{vanhorssen2016performance, vreman2022weaklyhardjl, vreman2022extended} (\hit{}, \miss{}, \recovery{}), or concatenate multiple outcomes using a \emph{lifted} model~\cite{linsenmayer2017stabilization, linsenmayer2021stabilization, seidel2024l2performance, seidel2026framework}.

\begin{myfigure}[t]
    \tikzexternaldisable
    \begin{tcolorbox}[colback=black!10!white, colframe=black, boxrule=1pt, fontupper=\small, boxsep=0pt, left=5pt, right=5pt, before skip=5pt, after skip=5pt]
    \textbf{Directed Graph}: 
    %a directed graph $\Graph$ is 
    a tuple $\langle \Nodes, \Edges \rangle$ where $\Nodes$ is a set of nodes and $\Edges$ is a set of labeled edges.
    Specifically, given an alphabet of labels $\Sigma$, we have $\Nodes= \{ v_1, \dots, v_{\numNodes} \}$ and $\Edges = \{ e_1, \dots, e_{\numEdges} \} \subseteq \Nodes \times \Sigma \times \Nodes$.
    An edge $e_k = (i,l,j) \in \Edges$ represents a connection from node $i \in \Nodes$ to node $j \in \Nodes$ labeled with $l \in \Sigma$.\\[2mm]
    \textbf{Markov Graph}: a directed weighted graph, i.e., a tuple $\langle \Nodes, \Edges \rangle$ where the label $l \in [0,1]$ of an edge $e_k = (i,l,j) \in \Edges$ is the transition probability from Markov mode $i$ to mode $j$.
    \end{tcolorbox}
    \caption{Directed and Markov graph definitions.}
    \label{box:graph}
    \vspace{-4mm}
\end{myfigure}

%%%%%%%%%%%%%%%%%%%%%%%%%%%%%%%%%%%%%%%%%%%%%%%%%%%%%%%%%%%%%%%%%%%%%%%%%%%%%%%%%%%%%%%%%%%%%%%%%%%%

\subsection{Controller classes}
\label{sec:basics:controller}
The control policy $\policy[\step]$ can be classified depending on its inputs.
Table~\ref{tab:policy-dependencies} shows the classification, rooted in control theory~\cite{aastrom2021feedback}, where different dependencies may be combined, as indicated using $(\cdot)$.

\begin{table}[ht]
    \centering
    \begin{threeparttable}
    \caption{Controller classes}
    \label{tab:policy-dependencies}
    \vspace{-0.3em}
    \rowcolors{1}{}{gray!10}
    \begin{tabular}{p{2cm} l p{1cm}}
        \toprule
        \textbf{Class} & \textbf{Controller input} & \textbf{Notation} \\
        \midrule
        State-feedback & System state $x[\step]$ & $\policy(x, \cdot)$ \\
        Output-feedback & System output $y[\step]$ & $\policy(y, \cdot)$ \\
        \midrule
        Dynamic & Controller state $z[\step]$ & $\policy(z,\cdot)$ \\
        Static  & -- & $\policy(\cdot)$ \\
        \midrule
        Switching & Pattern $(\mu[i])_{i \in \{\step-j-1, \dots, \step-1\}}$ & $\policy(\mu,\cdot)$ \\
        \midrule
        Time-dependent & $\step$ & $\policy(\step, \cdot)$ \\
        \bottomrule
    \end{tabular}
    \end{threeparttable}
\end{table}

The first distinction partitions state- and output-feedback controllers.
State-feedback controllers require to measure all system states (i.e., the vector $x[\step]$ at time step $\step$), while output feedback controllers rely on measuring the system output (i.e., the vector $y[\step]$ at time step $\step$).
The second distinction splits controllers into dynamic and static.
Dynamic controllers use internal information (such as estimates for physical quantities and integral of the error, stored in the controller state $z \in \mathbb{R}^{\dimCtrState}$), while static controllers do not possess an internal state.

A special type of internal state for switching controllers is the outcome of past iterations.
Some of the controllers compared in this paper use knowledge of a sequence of past outcomes $(\mu[i])_{i\in\{k-j-1,\dots,k-1\}}$ to adjust the controller calculations.
In Table~\ref{tab:policy-dependencies}, this sequence contains the last $j$ outcomes, which, for patterns of deadline hits and overruns, typically correspond to the most recent window $\window$.
For controllers designed using graph-based deadline miss models, adaptation to past outcomes is usually implemented by introducing a dependency of the control policy on the graph nodes encoding the current history of $\mu$. 
In the following, this is referred to as a node-dependent controller.

Finally, some controller may take time into account in the definition of the policy, for example by being more aggressive at the start and less aggressive as time progresses.

For each of the controllers analyzed in the following section, we will indicate to what class the controller belongs, and hence what set of information is needed for its implementation.
\section{Deadline-miss-aware controllers}
\label{sec:comparison}

We have selected eight methods for controller synthesis or adaptation that result in control systems that are deadline-miss aware, i.e.,~\cite{gaid2008dropin, vanhorssen2016performance, pazzaglia2019dmac, pazzaglia2021adaptive, vreman2022adaptive, gallant2024structure, gallant2025soft, seidel2026framework}.
In principle, other methods could also be adapted\footnote{Adapting original work meant to deal with packet dropouts or hardware faults would require modifying the proposed approaches with explicit design choices from our side.
We encourage the original proponents of these strategies to test such adaptations with our framework of Sec.~\ref{sec:experiments:framework}.} to the same setting, especially the ones dealing with packet dropouts~\cite{linsenmayer2021stabilization} in \gls{ncs} or hardware faults~\cite{ghosh2018faulttolerant}, but we restrict our study to the handling of deadline overruns.
In this section, we analyze the different characteristics of the controllers and provide brief descriptions of each design method.

Table~\ref{tab:comparison} reports the distinguishing features of each controller.
We partition these features into three categories:
\begin{enumerate}[label=\alph*)]
    \item how the method models and handles deadline overruns;
    \item controller characteristics; and
    \item implementation aspects.
\end{enumerate}

In Table~\ref{tab:comparison}$\,$(a), we first classify the modeling and handling of deadline misses as stochastic (\iconStochastic{}) or deterministic (\iconDeterministic{}) approaches.
We then indicate the specific deadline-miss models, including if the method can be generalized by extending it to consider any regular language (see Sec.~\ref{sec:basics:missmodel}).
Finally, we report on the supported deadline overrun and actuator strategies (see Sec.~\ref{sec:basics:misses}), which are key design choices that can impact both applicability and performance.

\begin{myfigure}[t]
    \tikzexternaldisable
    \begin{tcolorbox}[colback=black!10!white,  colframe=black, boxrule=1pt, fontupper=\small, boxsep=0pt, left=5pt, right=5pt, before skip=5pt, after skip=5pt]
    \textbf{Quadratic cost} $J(x,u) = \Sigma_{\step=0}^{N-1}x[\step]^\top Q_\mathrm{x} x[\step]+u[\step]^\top Q_\mathrm{u} u[\step] + x[N]^\top Q_\mathrm{N} x[N] $: weighted sum of states and control action over a finite or infinite (if $N=\infty$, $Q_\mathrm{N}=0$) time horizon, combining state convergence and control effort, with positive definite $Q_\mathrm{u} \in \mathbb{R}^{\dimInput \times \dimInput}$ and positive semidefinite $Q_\mathrm{N},Q_\mathrm{x} \in \mathbb{R}^{\dimState \times \dimState}$. \\[1.5mm]
    \textbf{Expected quadratic cost} $\mathbb{E}[J(x,u)]$: a probabilistic version of the quadratic cost averaged over the possible hit-miss patterns.\\[1.5mm]
    \textbf{$\ell_2$-performance}: measures the worst-case energy amplification from an (artificial) exogenous signal $\bar{u}$, representing, e.g., disturbances, to an performance output $\bar{y} = \bar{C}x[\step] + \bar{D}u[\step]$, representing quantities of interest, e.g, costs.
    The $\ell_2$-gain is the smallest $\gamma$ such that $\Sigma_{\step=0}^{\infty} \bar{y}[\step]^\top \bar{y}[\step] < \gamma^2 \Sigma_{\step=0}^{\infty} \bar{u}[\step]^\top \bar{u}[\step]$ holds.
    \end{tcolorbox}
    \caption{Cost function definitions.}
    \label{box:cost-functions}
    \vspace{-4mm}
\end{myfigure}

\begin{myfigure}[t]
    \tikzexternaldisable
    \begin{tcolorbox}[colback=black!10!white,  colframe=black, boxrule=1pt, fontupper=\small, boxsep=0pt, left=5pt, right=5pt, before skip=5pt, after skip=5pt]
    Stability is defined w.r.t.\ an equilibrium point of the system, typically $0$.\\[1.5mm]
    \textbf{Asymptotic stability}: if the system's state starts close to $0$, it stays close to $0$ for all times and the state $x[\step]$ converges asymptotically to $0$ (i.e., $x[\step] \rightarrow 0$).
    This is often verified using spectral radius arguments, or using Lyapunov methods.
    \\[1.5mm]
    \textbf{Mean-square stability}: the second moment of the state $x[\step]$ converges to $0$ (i.e., $\mathbb{E}\{x[\step]^T x[\step]\} \rightarrow 0$), indicating the stochastic convergence of $x[\step]$ to $0$.
    \end{tcolorbox}
    \caption{Stability definitions.}
    \label{box:stability}
    \vspace{-4mm}
\end{myfigure}

In Table~\ref{tab:comparison}$\,$(b), we outline the choices concerning the control design.
These include the selection of the control policy class (Sec.~\ref{sec:basics:controller}), the theoretical guarantees offered by the method, and the targeted performance criteria, i.e., the cost function used in the control-design optimization problem.
We outline the cost functions relevant to this paper in Box~\ref{box:cost-functions}.
The theoretical guarantees can concern either the stability (see Box~\ref{box:stability}) of the closed-loop system, or control performance metrics, including constraints on states and inputs.

In Table~\ref{tab:comparison}$\,$(c), we consider the implementation aspects:
This includes whether the control policy needs to know the deadline-miss pattern $\mu$, the computational load (offline for the design, and online for the execution of the policy $\policy$), and the control policy's memory footprint.
The offline part is run once at design time, while the online part consists of the computations done in every control job.
The memory footprint influences the storage requirements of the controller task.

\begingroup
\begin{table*}[ht]
    \tikzexternaldisable
    \centering
    \setlength{\abovetopsep}{-1ex}
    \renewcommand{\arraystretch}{1.15}
    \caption{Control method comparison: overview} \label{tab:comparison}
    
\begin{subtable}{\textwidth}\begin{threeparttable}
    \caption{Deadline-miss model and real-time setting}
    \label{tab:comparison-constraints}
    \rowcolors{1}{}{gray!10}
    \begin{tabular}{p{4.3cm}cp{3.7cm}p{1.7cm}p{3.05cm}p{1.55cm}}
        \toprule 
         \textbf{Method (authors, year)} & \multicolumn{2}{c}{\textbf{Deadline-miss model}} & \makecell[c]{\textbf{Extendable to}\\\textbf{regular lang.}} & \makecell[c]{\textbf{Overrun strategy} \stratov} & \makecell[c]{\textbf{Actuator}\\\textbf{strategy} \stratac} \\
         \midrule
         \paperNameGaid{} (Gaid et al.{}\,'08) \cite{gaid2008dropin} & \scalebox{0.8}{\iconDeterministic} & \anyHit{\nhit}{\window} with specific pre-determined periodic hit pattern & \centering \no & \Kill & \phantom{\Zero,} \Hold \\
         \paperNameHorssen{} (van Horssen et al.{}\,'16) \cite{vanhorssen2016performance} & \scalebox{0.8}{\iconDeterministic} & \anyHit{\nhit}{\window} & \centering \yes & \Kill & \parbox[t]{2.1cm}{\Zero, (\Hold)} \\
         \paperNameDMAC{} (Pazzaglia et al.{}\,'19) \cite{pazzaglia2019dmac} & \scalebox{0.8}{\iconStochastic} & \multicolumn{2}{l}{n.a.~- based on example schedules} & \Kill, \Skip, \Queue & {\Zero,} \Hold \\
        \paperNameStreamAdapt{} (Pazzaglia et al.{}\,'21) \cite{pazzaglia2021adaptive}& \scalebox{0.8}{\iconDeterministic} & adaptable to \rowMiss{\nmiss} & \centering \no & \phantom{\Kill,} \Skip & \phantom{\Zero,} \Hold \\
         \paperNameVreman{} (Vreman et al.{}\,'22) \cite{vreman2022adaptive} & \scalebox{0.8}{\iconDeterministic} & \rowMiss{\nmiss} & \centering \no & \Kill & \Zero, \Hold \\
         \paperNameMJLS{} (Gallant et al.{}\,'24) \cite{gallant2024structure} & \scalebox{0.8}{\iconStochastic \hspace{0.1mm} / \iconDeterministic} & $\sim$ \rowMiss{1} & \centering \yes & \centering Hybrid strategy\tnote{a} & \phantom{\Zero,} \Hold\tnote{a} \\
         \paperNameMPC{} (Gallant et al.{}\,'25) \cite{gallant2025soft} & \scalebox{0.8}{\iconStochastic} & $\rowMiss{\nmiss}$ & \centering \yes\tnote{b} & \Kill, \Skip, \Queue & \Zero, \Hold \\
         \paperNameMS{} (Seidel et al.{}\,'26) \cite{seidel2026framework} & \scalebox{0.8}{\iconDeterministic} & \multicolumn{2}{c}{naturally: any given by a regular language\tnote{c}} & \Kill, \Skip & \Zero, \Hold \\
         \bottomrule
    \end{tabular}
\end{threeparttable}\end{subtable}

\vspace{0.5em}

\begin{subtable}{\textwidth}
    \caption{Controller design}
    \label{tab:comparison-design}
    \rowcolors{1}{}{gray!10}
    \begin{tabular}{llp{5.75cm}p{5.75cm}}
         \toprule
         \textbf{Method}  & \textbf{Controller class} & \textbf{Performance criteria for design} & \textbf{Theoretical guarantees}\\
         \midrule
         \paperNameGaid{} \cite{gaid2008dropin} & $\policy(\step,x)$  & general performance metric;  e.g.,~quadratic costs for \acs{lti} systems & retains guarantees of controller designed for periodic hit pattern, more hits improve performance \\
         \paperNameHorssen{} \cite{vanhorssen2016performance} & $\policy(x,\mu)$ & infinite-horizon quadratic cost & asympt. stability and quadratic costs upper bound \\
         \paperNameDMAC{} \cite{pazzaglia2019dmac} & $\policy(x)$  & expected quadratic cost term & none \\
         \paperNameStreamAdapt{} \cite{pazzaglia2021adaptive} & $\policy(y,\mu)$ or $\policy(y,\mu,z)$  & general performance metric
         & a-posteriori stability check based on \acs{jsr} analysis \\
         \paperNameVreman{} \cite{vreman2022adaptive} & $\policy(y,\mu)$ or $\policy(y,\mu,z)$  & n.a.~(adapts a given controller) & a-posteriori stability and performance analysis \\
         \paperNameMJLS{} \cite{gallant2024structure} & $\policy(x,\mu)$ or $\policy(x)$  & criteria depend on objective function (e.g., worst-case expected infinite-horizon quadratic cost) & mean-square stability with confidence $1-\beta$\\
         \paperNameMPC{} \cite{gallant2025soft} & $\policy(x,\mu)$  & expected quadratic cost term of states and inputs and minimization of soft state constraints violation  & mean-square stability, recursive feasibility, satisfaction of input/state constraints \\
         \paperNameMS{} \cite{seidel2026framework} & $\policy(x,\mu)$ or $\policy(x)$  & $\ell_2$-performance (worst-case energy amplification) & asymptotic stability and $\ell_2$-performance \\
         \bottomrule
    \end{tabular}
\end{subtable}

\vspace{0.5em}

\begin{subtable}{\textwidth}\begin{threeparttable}
    \caption{Implementation aspects}
    \label{tab:comparison-implementation}
    \rowcolors{1}{}{gray!10}
    \begin{tabular}{lcp{3.2cm}p{3.9cm}p{4.5cm}}
        \toprule
        \textbf{Method} & \textbf{Miss-knowledge} &  \textbf{Comput. load: Offline} &  \textbf{Comput. load: Online}\tnote{d} & \textbf{Memory footprint} \\
        \midrule
         \paperNameGaid{} \cite{gaid2008dropin} & \no & $w$ \acs{are}s (\acs{lti} systems) & \matvecmult{\dimInput}{\dimExtendedState}{\dimExtendedState} (\acs{lti} systems) & $w$ time-dependent gains\tnote{e} ~(\acs{lti} systems) \\
         \paperNameHorssen{} \cite{vanhorssen2016performance} & \yes &  \acs{sdp} or iterative \acs{sdp} design & \matvecmult{\dimInput}{\dimExtendedState}{\dimExtendedState} & node-dependent gains\tnote{e} \\
         \paperNameDMAC{} \cite{pazzaglia2019dmac} & \no & \acs{sre} & \matvecmult{\dimInput}{\dimExtendedState}{\dimExtendedState} & node-independent gain\tnote{e} \\
         \paperNameStreamAdapt{} \cite{pazzaglia2021adaptive} & \yes & $\numControllers\,\times$ SDP or classic tuning  & \matvecmult{\dimInput}{\dimExtendedState}{\dimExtendedState} & depends on \acs{wcrt} and granularity \\
         \paperNameVreman{} \cite{vreman2022adaptive} & \yes & simple matrix computations & \matvecmult{(\dimCtrState + \dimInput)}{(\dimCtrState + 2\dimOutput)}{\dimCtrState + 2\dimOutput} & node-dependent dynamic controller\tnote{e} \\
         \paperNameMJLS{} \cite{gallant2024structure} & \yes or \no & \acs{sdp} & \matvecmult{\dimInput}{(\dimState + 2 \dimInput)}{\dimState + 2 \dimInput} & node-dependent or -independent gains\tnote{e} \\ 
         \paperNameMPC{} \cite{gallant2025soft} & \yes &
         \acs{sdp} & \acs{socp} & node-dependent OCPs \\ 
         \paperNameMS{} \cite{seidel2026framework} & \yes or \no & \acs{sdp} & \matvecmult{\dimInput}{\dimExtendedState}{\dimExtendedState} & node-dependent or -independent gains\tnote{e} \\
         \bottomrule
    \end{tabular}
    \begin{tablenotes}
        \item $^\text{a}$ Extendable to any overrun strategy $\stratov \in \stratovset$ and actuator strategy $\stratac \in \stratacset$.
        $\quad^\text{b}$ Probabilities are required.
        $\quad^\text{c}$ Includes all WH constraints.\\
        $\quad^\text{d}$ Matrix-vector multiplication is denoted as \matvecmultOrig{}{}, with respective dimensions. Control methods with miss-knowledge require also the current node.\\
        $\quad^\text{e}$ The number of stored coefficients (per node-dependent gain) deduces from the size of the matrices in the previous column.
    \end{tablenotes}
\end{threeparttable}\end{subtable}

\end{table*}
\endgroup

In the following, the compared controller synthesis methods are briefly described.

%%%%%%%%%%%%%%%%%%%%%%%%%%%%%%%%%%%%%%%%%%%%%%%%%%%%%%%%%%%%%%%%%%%%%%%%%%%%%%%%%%%%%%%%%%%%%%%%%%%%

\subsection{\paperNameGaid{} \cite{gaid2008dropin}}
\label{sec:controllers:acccontrol}

This controller synthesis method uses an \anyHit{h}{w} model under \Kill{} and \Hold, including an additional assumption that some deadlines corresponding to a worst-case execution pattern are always met.
Following this special model, \emph{mandatory} and \emph{optional control jobs} are defined.
Jobs that must hit their deadline in order to fulfill the constraint are classified as mandatory, while all other jobs are optional.
The periodicity of the mandatory jobs' pattern influences scheduling decisions.

The model allows to introduce the notion of \emph{accelerable control} (\paperNameGaid{}):
A control law is accelerable if any deadline hit of an optional job improves the control performance.
The controller is then designed based on the worst-case known hit pattern, i.e., assuming that only the deadline of the mandatory jobs hit their deadlines.
Using the Bellman optimality principle yields an optimization-based control design method, in which this controller is refined to make it accelerable.
This method is compatible with the \gls{let} paradigm using state augmentation.
While theoretically this synthesis method can work for nonlinear systems and for any general performance index, the resulting problem is computationally tractable only for certain classes of systems.
For \gls{lti} systems, the work specifically designs a controller for quadratic costs, solving \glspl{are}.

\subsection{\paperNameHorssen{} \cite{vanhorssen2016performance}}
\label{sec:controllers:graphlqr}

This control synthesis method assumes an \anyHit{\nhit}{\window} deadline-miss model, and models the closed-loop system as a switched system with constrained switching.
This is one of the first controller synthesis methods that introduces a graph to describe the admissible patterns of deadline hits and misses.
In \paperNameHorssen{}, state-feedback controllers are designed using a \gls{lqr} design method with an infinite horizon quadratic cost function.
The method supports the \Kill{} and \Zero{} strategies, but can be extended to \Hold{}.

The approach is a two-step procedure.
First, a standard \gls{lqr} for the case without deadline misses is analyzed with respect to its worst-case performance loss under the deadline-miss model.
The analysis can be performed using \gls{lmi}-based tools, resulting in solving a \gls{sdp}, or using a more conservative analytic upper bound.
Second, a node-dependent controller is designed based on the graph.
The design is based on either solving an \gls{sdp} or on using a computationally less expensive (but more conservative) iterative design approach that exploits the \gls{lqr} from the first step.
The method guarantees asymptotic stability and an upper bound on the worst-case quadratic cost.

\subsection{\paperNameDMAC{} \cite{pazzaglia2019dmac}}
\label{sec:controllers:dmac}

This controller synthesis method uses a stochastic model for deadline misses and an \gls{lti} plant model.
The approach considers the reactive overrun strategies, \Kill{}, \Skip{}, and \Queue. It uses \Hold{} as actuation strategy, but an extension to \Zero{} appeared in~\cite{pazzaglia2020performance}.
The approach requires performing a certain number of simulations for the scheduling of the entire task set to obtain the worst-case probabilities of specific patterns of deadline overruns, and identifying the switched dynamics equations under the different strategies.

To provide robustness guarantees on the estimation of the overrun probabilities,
the approach uses \emph{scenario theory}~\cite{calafiore2006scenario}.
This ensures that the most pessimistic sequence selected in a set of simulations can be classified as close to the real ``worst'' sequence under given robustness and confidence levels.
Based on the sequence, a fixed \gls{lqg} controller is obtained by solving the corresponding \gls{sre}, paired with a linear predictor.
The resulting controller is shown to have better performance than a controller with larger period without overruns
and \Skip{} is shown to obtain better performance on average with respect to the other two strategies for small control periods (i.e., in case of higher probabilities of missing deadlines).

\subsection{\paperNameStreamAdapt{} \cite{pazzaglia2021adaptive}}
\label{sec:controllers:streamadapt}

This work targets control applications where overruns may happen sporadically, and the knowledge of the timing model is limited to the \gls{wcrt} $\wcrt$ of the control task (and thus adaptable to the \rowMiss{\nmiss} model).
The underlying control design is a standard periodic one-step-delayed feedback controller with actuator handling \Hold{}, and with period~$\period$ being either a multiple of (or equal to) the sensor period~$\periodSensor$.
The proposed method extends this basic controller with an adaptive design based on the \emph{continuous stream} model of computation~\cite{fontanelli2013continuous}.
Here, a job missing its deadline is allowed to continue executing by ``stretching'' its periodic instance by multiples of the sensor period until completion, while postponing the activation of the next control job.
The next job then re-initializes the periodic pattern and can effectively execute without any interference from the previous job. 
For the case of $\period=\periodSensor$, such overrun strategy coincides with \Skip.
To compensate the deviation in the expected state evolution during an overrun, the job that is activated after an overrunning one is executed with a different controller, tuned as a function of the length of the overrun (quantized by the number of sensing periods $\periodSensor$). 
Under the assumption that the \gls{wcrt} is known, a finite number of control tunings is thus required, equal to $\numControllers = \lceil(\wcrt-\period)/\periodSensor\rceil$.
Asymptotic stability is evaluated in terms of the \gls{jsr} and the approach is compared against other non-adaptive designs under different \gls{wcrt} conditions, showing increased flexibility and performance.

\subsection{\paperNameVreman{} \cite{vreman2022adaptive}}
\label{sec:controllers:dmai}

\paperNameVreman{} is an implementation method for dynamic \gls{lti} controllers to improve the system performance and stability under deadline misses, that does not affect the system's performance in their absence.
While the controller adaptation is evaluated only for the \Hold{} case, the approach is independent of the actuator strategy.
However, it requires the \Kill{} overrun strategy, and knowledge, during execution, of the number of deadlines that have been missed since the last hit (using \rowMiss{\nmiss}).
When one or more deadline misses occur, the proposed approach updates the controller state with the value that it would have had if the deadline misses did not occur.
This computation is performed using the controller equations and propagating its state over time based on the number of missed task executions.
The missing sensor readings of the killed jobs are approximated with a linear interpolation between the reading from last hit and the newest one.
This translates in making the \gls{lti} controller matrices a function of the number of consecutive deadlines that have been missed.

The approach requires a set of six matrices for each number of possible consecutive deadline misses, which can be computed offline and stored to avoid increasing the online computational cost.
The size of each matrix is upper bounded by the size \dimCtrState{} of the controller's state.
While the paper does not provide theoretical guarantees, it includes a probabilistic analysis of the performance and a \gls{jsr}-based analysis of the stability showing a significant improvement against the non-adaptive controller for many industrial plants.

\subsection{\paperNameMJLS{} \cite{gallant2024structure}}
\label{sec:controllers:drmjls}

\paperNameMJLS{} is a synthesis method that finds a \textit{distributionally robust} state feedback control method for stabilizing discrete-time \textit{\glspl{mjls}}.
A real-time control system subject to bounded probabilistic deadline overruns can be modeled as such a switching system, by encoding distinct parts of hit/miss patterns as Markov modes with uncertain transition probabilities.
An example is provided, where a hybrid overrun strategy incorporating both \Kill{} and \Queue{(1)} in combination with \Hold{} is used. 
The example section briefly motivates different sources of uncertainty regarding the probabilities of missing a deadline that are compensated for in the controller design. 
The approach guarantees (with a given confidence) mean-square stability for \glspl{mjls}, where the transition probabilities can be time-varying and/or are estimated based on a finite number of samples.
Depending on the available data, the resulting controller can be robust over the full space of probabilities (i.e., deterministic) or for a limited set of possible probabilities, converging for a large number of samples to a fully stochastic one.
While the paper shows only one specific example with \rowMiss{1}, the extension to other deadline-miss models, overrun, and actuator strategies is straightforward based on the steps used to model the included example system.
Either a node-dependent or a non-switching state-feedback controller is designed offline based on distributionally robust Lyapunov-type conditions transformed into sets of \glspl{lmi}.
The offline design scales with $\bigO{\dimExtendedState^{6.5} \ln{\dimExtendedState}}$ and $\bigO{\numNodes^{3.5} \ln{\numNodes}}$.

\subsection{\paperNameMPC{} \cite{gallant2025soft}}
\label{sec:controllers:treempc}

\paperNameMPC{} is a stochastic \gls{mpc} approach for \glspl{mjls}. The method guarantees mean-square stability by optimizing a finite horizon expected quadratic cost, given the transition probabilities.
Additionally, it provides robust satisfaction of any hard input/state constraints, and penalizes violations of soft constraints.
The original formulation encodes \rowMiss{\nmiss} but can be extended to any regular language, if a probabilistic modelization is available.
The work demonstrates implementations for the overrun strategies \Kill{}, \Skip{}, and \Queue{(1)}, combined with the actuator overrun models \Zero{} and \Hold{}.
Based on a \gls{mjls} model, the controller is designed by spanning a scenario tree of potential future patterns of deadline outcomes from each relevant initial mode and compiling \glspl{ocp} using the predicted state trajectories for these trees.

To guarantee recursive feasibility of the online optimization problem, the offline design step synthesises a mode-dependent terminal set and cost function by solving a set of \glspl{lmi} that scale with $\bigO{\dimExtendedState^{6.5} \ln{\dimExtendedState}}$ and $\bigO{\numNodes^{3.5} \ln{\numNodes}}$.
The online phase of a control job involves compiling the augmented state vector using state measurements and auxiliary variables, then solving an \gls{ocp} for the current initial mode.
Due to the scenario-tree formulation, these \glspl{ocp} can be formulated as computationally tractable \glspl{socp} and solved efficiently using interior point methods.

\subsection{\paperNameMS{} \cite{seidel2026framework}}
\label{sec:controllers:gss}

The graph-based switched system controller design approach (\paperNameMS{}) is an \gls{lmi}-based synthesis method that explicitly incorporates the deadline-miss model in the design phase.
It considers all evolutions of the system under all possible patterns of deadline outcomes.
For all the patterns, it provides rigorous guarantees for asymptotic stability and $\ell_2$-performance, even in the worst-case.
Analyzing a given controller for its achieved performance is also possible.
This is thus a deterministic approach, that can handle \Zero{} and \Hold{}, and the two overrun strategies \Kill{} and \Skip{}.
The design part is done offline: 
A switched system description of the \gls{wh} control system together with the concept of multiple Lyapunov functions is used to formulate a set of \glspl{lmi}, which result in a computationally tractable \glspl{sdp}.
Its computational effort scales with $\bigO{\dimExtendedState^{6.5} \ln{\dimExtendedState} }$ and $\bigO{\numNodes^{3} \numEdges^{0.5} \ln{\numEdges} + \numNodes^{2} \numEdges^{1.5} \ln{\numEdges} }$.
Note that the (lifted) graph used in this work is smaller than the one used in other graph-based methods.

The resulting controller additionally utilizes the last applied input, which improves performance.
It is possible to either design multiple state-feedback controller gains, from which at start of each job one is picked depending on the past hit/miss-pattern (resulting in a node-dependent controller), or to design a single controller gain valid for  all admissible hit/miss-patterns, although more conservative.
The only requirement on the deadline-miss model is that it can be represented by a labeled directed graph, making it thus naturally applicable to any \gls{wh} constraint, but also to any deadline-miss model that corresponds to a regular language.

\section{Qualitative comparison}
\label{sec:qualitativecomparison}
Based on Table~\ref{tab:comparison}, we qualitatively compare the considered deadline-miss-aware controllers along two main dimensions: \emph{applicability}, and \emph{control design criteria and guarantees}.
Our objective is to expose the trade-offs that appear in the state-of-the-art controllers with deadline-miss awareness.
Such trade-offs can guide practitioners in the choice of a suitable controller for their applications based on
\begin{enumerate*}[label=(\roman*)]
    \item the  assumptions about the system, the expected implementation effort and available computational resources, and
    \item the desired control performance and guarantees.
\end{enumerate*}
At the same time, researchers can use this discussion to identify open research directions.

\subsection{Applicability}
\label{sec:qualitative-applicability}
Applicability mainly depends on three aspects: model assumptions on the real-time system, assumptions on the control design setting, and implementation effort offline and online.

\smallskip
\emph{Underlying assumptions about the real-time system.}
The methods differ significantly in the assumed structure for deadline hit/miss patterns.
\paperNameGaid{} is the most restrictive, since it relies on a predetermined periodic pattern of mandatory hits fulfilling an \anyHit{\nhit}{\window}-like constraint.
This is attractive when such a pattern can be enforced, but less suitable when misses are irregular or only partially characterized.
At the other extreme, \paperNameVreman{} and \paperNameStreamAdapt{} require only limited timing knowledge:
\paperNameVreman{} needs the number of consecutive misses since the last successful execution, while \paperNameStreamAdapt{} requires only the task's \gls{wcrt} and adapts the controller to the overrun length.
These methods are therefore easier to deploy when no precise miss model is available.

A middle ground is provided by methods that explicitly encode admissible miss patterns. 
\paperNameHorssen{} uses an $(h,w)$-firm model but can be extended to regular languages, while \paperNameMS{} handles any deadline-miss model representable as a labeled graph, including \gls{wh} constraints and regular languages.
\paperNameMPC{} and \paperNameMJLS{} are also flexible, since their \gls{mjls} formulation can encode different hit/miss histories and execution semantics, provided the corresponding modes and transitions are available.
In practice, the closer this model is to the actual platform behavior, the less conservative the synthesis becomes.

A second applicability aspect are the supported combinations of overrun and actuator strategies.
As shown in Table~\ref{tab:comparison}$\,$(a), several methods are tied to \Kill{}, while fewer also support \Skip{}.
Likewise, some methods are derived only for \Hold{} or only for \Zero{}, whereas \paperNameDMAC{}, \paperNameMPC{}, \paperNameMJLS{}, and \paperNameMS{} cover a broader set of combinations.
This is an important distinction because the reaction to overruns directly changes the closed-loop dynamics, as discussed in Sec.~\ref{sec:basics:misses}.
Hence, a theoretically appealing method may still be unusable if the runtime semantics do not match.

\iffalse
\begin{table}[t]
    \centering
    \caption{Controller applicability}
    \label{tab:controllers-applicability}
    \begin{tabular}{p{0.75cm}ll}
        \toprule
                         & \makecell[c]{\textbf{\Kill{}}}               & \makecell[c]{\textbf{\Skip{}}} \\
        \midrule
        \rowcolor{gray!10}
        &
            \paperNameHorssen{},
            \paperNameDMAC{},
            \paperNameVreman{}, 
        &
            \paperNameDMAC{},
            \paperNameMJLS{}, 
            \paperNameMPC{}, 
        \\
        \rowcolor{gray!10}
        \multirow{-2}{*}{\textbf{\Zero{}}}
            &
            \paperNameMJLS{}, 
            \paperNameMPC{}, 
            \paperNameMS{}
            &
            \paperNameMS{} 
            \\
        \multirow{2}{*}{\textbf{\Hold{}}}
        & 
            \paperNameGaid{},
            \paperNameDMAC{},
            \paperNameVreman{},
        &
            \paperNameDMAC{},
            \paperNameStreamAdapt{}
            \paperNameMJLS{}, 
            \\
        &
            \paperNameMJLS{}, 
            \paperNameMPC{},
            \paperNameMS{}
        &
            \paperNameMPC{},
            \paperNameMS{}
            \\
        \bottomrule
    \end{tabular}
\end{table}
\fi

Most approaches also assume availability of online information about recent deadline outcomes.
This introduces integration effort and state bookkeeping.
Sequence-aware controllers such as \paperNameHorssen{}, \paperNameMPC{}, and \paperNameMS{} use this information to adapt gains or optimization problems to the current timing mode, reducing conservatism at the cost of higher complexity.
By contrast, fixed-gain approaches such as \paperNameDMAC{} reduce runtime dependence on miss-history tracking, but forfeit the added degree of freedom through adaptation.

\smallskip
\emph{Underlying assumptions for the control design.}
The second column of Table~\ref{tab:comparison}$\,$(b) shows that the most common controller structure is sequence-aware static state-feedback.
This is the case for \paperNameHorssen{} and the node-dependent versions of \paperNameMJLS{} and \paperNameMS{}.
Similarly, \paperNameMPC{} proposes a non-static sequence-aware state feedback control policy and \paperNameGaid{} also uses state feedback, but with explicit time dependence induced by the mandatory/optional job pattern.
These methods are attractive when full state measurement is available or when an observer can be incorporated.
However, this may be restrictive when only outputs are measured.
\paperNameStreamAdapt{} and \paperNameVreman{} are advantageous here, since they support output-feedback and dynamic controllers and can therefore be integrated into a broader class of existing industrial designs.

Another distinction is whether a method synthesizes a controller from scratch or adapts an existing one.
\paperNameHorssen{}, \paperNameGaid{}, \paperNameMJLS{}, \paperNameMPC{}, and \paperNameMS{} are synthesis frameworks requiring a plant model and producing a dedicated control law.
By contrast, \paperNameVreman{} is an implementation-level adaptation of a given controller, which is attractive when redesign is undesirable.
\paperNameStreamAdapt{} lies in between, starting from a periodic controller and augmenting it with a finite set of retuned variants.
Thus, \paperNameVreman{} and \paperNameStreamAdapt{} are more relevant in legacy settings to robustify existing controllers.

Finally, most methods assume an \gls{lti} plant model, or are mainly tractable in that setting.
Nonlinear systems are only marginally addressed, which reflects a broader limitation of the current state of the art.
While there has been some work on the analysis of nonlinear systems under \gls{wh} constraints~\cite{hertneck2020nonlinear, hertneck2021efficient}, the controller design for those systems, even for certain system classes, is an interesting topic for future research.
Note that \paperNameGaid{} proposes a design method for nonlinear plants that is based on Bellman's optimality principle, which is in general hard to solve for nonlinear dynamics.

\smallskip
\emph{Implementation effort.}
Offline, optimization-based synthesis is the dominant paradigm.
\paperNameHorssen{}, \paperNameMJLS{}, \paperNameMS{}, and \paperNameMPC{} (for terminal ingredients) rely on convex programs such as \glspl{sdp}, while \paperNameMPC{} additionally requires the construction of scenario-trees and corresponding \glspl{ocp}.
This yields strong guarantees, but also a nontrivial synthesis pipeline.
\paperNameDMAC{} instead solves an \gls{sre} after estimating a probabilistic deadline model from scheduling simulations, shifting part of the effort to timing-data generation and calibration.
\paperNameGaid{} is lighter in the \gls{lti} quadratic-cost case, where the design reduces to \glspl{are}, but this comes with stronger assumptions on the hit/miss pattern.
\paperNameVreman{} has the smallest redesign burden, requiring only simple matrix computations from an existing controller, while \paperNameStreamAdapt{} can be implemented either through repeated tuning or by solving a bounded number of \glspl{sdp} depending on the \gls{wcrt} granularity.

More broadly, some methods require not only solving the control problem, but also obtaining the deadline-miss model itself.
For deterministic graph-based methods, this means identifying a regular-language or \gls{wh} abstraction of admissible patterns; for stochastic methods, it means estimating transition probabilities or representative schedules.
In practice, this modeling step can dominate the engineering effort.
Accordingly, methods with weaker timing-model requirements may be easier to adopt even if they are less specialized theoretically.

\smallskip
\emph{Runtime computation overhead.}
Most methods have light runtime load.
\paperNameGaid{}, \paperNameHorssen{}, \paperNameDMAC{}, \paperNameMJLS{}, and \paperNameMS{} mainly reduce to selecting a gain matrix, possibly depending on the current timing mode, and performing a matrix-vector multiplication.
Their runtime burden is therefore modest and predictable, although sequence-aware or node-dependent controllers require storing multiple controller gains.

\paperNameVreman{} adds state-update computations for the adapted dynamic controller, but these remain simple linear-algebra operations.
\paperNameStreamAdapt{} selects among a finite number of retuned controllers according to the observed overrun length, so its overhead scales with the discretization granularity.
\paperNameMPC{} is the clear online outlier, since it solves an optimization problem during execution.
Even if the resulting \gls{socp} is tractable, this remains substantially more demanding and may limit implementation on constrained platforms.
Thus, \paperNameMPC{} offers the greatest flexibility in handling constraints and stochastic evolution, but at the highest runtime computational cost.

Overall, applicability spans a spectrum.
At one end are methods such as \paperNameVreman{} and \paperNameStreamAdapt{}, which are easier to integrate but rely on weaker timing models and more limited adaptation mechanisms.
Their closed-loop behaviors depend on the baseline controllers that are not necessarily designed for good performance of the deadline-miss-aware controllers.
At the other end are methods such as \paperNameMS{}, \paperNameMJLS{}, and \paperNameMPC{}, which explicitly encode the timing model and exploit it effectively, but require more modeling effort and, in some cases, significantly higher computational resources.

\subsection{Control design criteria and guarantees}
\label{sec:qualitative-guarantees}

The second comparison dimension concerns what the methods optimize and what they can guarantee.
We first discuss the design criterion and then the closed-loop guarantees.

\smallskip
\emph{Performance criteria for design.}
Criteria relating to quadratic cost terms (Box~\ref{box:cost-functions}) dominate the literature, with
\paperNameGaid{}, \paperNameHorssen{}, \paperNameDMAC{}, and \paperNameMPC{} all optimizing or analyzing variants of this criterion.
This is natural, since quadratic objectives fit well with \gls{lti} models and lead to tractable formulations.
\paperNameMPC{} extends this by combining the finite-horizon expected quadratic cost with explicit handling of soft state constraints.
\paperNameMS{}, by contrast, uses an $\ell_2$-performance objective (see Box~\ref{box:cost-functions}), 
which is more aligned with robust analysis than average-cost optimality.
\paperNameVreman{} does not define a new design criterion, since it adapts a given controller and is largely agnostic to its original objective.

Crucially, the performance criterion maps directly to the targeted operating regime, favoring nominal efficiency, robustness, or constraint handling.
When average behavior under stochastic misses is of primary interest, expected quadratic cost formulations are well suited.
Conversely, when miss patterns must be tolerated adversarially within a deterministic admissible set, worst-case criteria as in \paperNameMS{} and \paperNameHorssen{} or the worst-case analysis in \paperNameStreamAdapt{} are more appropriate.
In other cases, minimal implementation overhead is the priority, favoring an off-the-shelf approach like \paperNameVreman{}.

\smallskip
\emph{Stability guarantees.}
Most methods provide some form of stability guarantee (Box~\ref{box:stability}), depending on the framework.

For deterministic switched system formulations such as \paperNameHorssen{} and \paperNameMS{}, the guarantee is \textit{asymptotic stability} under all admissible switching sequences.
These are strong guarantees, in that they cover all timing patterns allowed by the model. 
\paperNameStreamAdapt{} does not guarantee stability by design, but provides a framework for a-posteriori stability, relying on a \gls{jsr}-based analysis over admissible overrun evolutions.
Similarly, \paperNameVreman{} uses both a time-averaged performance analysis and a \gls{jsr}-based analysis to analyze the performance and stability of the resulting system.

For \gls{mjls}-based stochastic approaches, the relevant notion is \textit{mean-square stability}.
\paperNameMPC{} guarantees this together with recursive feasibility.
\paperNameMJLS{} guarantees mean-square stability, but in a distributionally robust sense: the guarantee holds with a certain confidence even when transition probabilities are uncertain or estimated from finite samples, making it attractive when stochastic timing behavior is not exactly known.
\paperNameDMAC{}, instead, is mainly motivated by performance under estimated stochastic miss patterns and does not provide an analogous formal stability certificate.

Taken together, these guarantees reflect the broader positioning of the methods.
Graph-based deterministic methods offer strong worst-case statements under explicit admissible languages of deadline outcomes.
\gls{mjls}-based methods trade such adversarial coverage for stochastic notions that better exploit probability information.
Adaptation methods such as \paperNameVreman{} and, to a lesser extent, \paperNameStreamAdapt{} emphasize practical improvement under overruns, with lighter integration effort but generally weaker or narrower guarantees.

\smallskip
\emph{Additional guarantees.}
Only a subset of the methods provides guarantees beyond stability.
\paperNameHorssen{} gives an upper bound on the worst-case quadratic cost over all admissible hit/miss patterns under its $(h,w)$-firm model.
\paperNameMS{} guarantees an $\ell_2$-performance level in the worst case.
\paperNameMPC{} is the richest in this respect: besides mean-square stability, it guarantees recursive feasibility and satisfaction of hard input/state constraints while minimizing expected cost and penalizing soft-constraint violations.
\paperNameGaid{} provides a different but also meaningful guarantee: it preserves the guarantees of the controller designed for the mandatory hit-pattern, and any additional optional hit improves performance.
By contrast, \paperNameDMAC{} and \paperNameVreman{} mainly show empirical or comparative improvements without formal performance bounds.
This highlights a recurring trade-off: stronger guarantees usually require more structured models and heavier synthesis, whereas lighter assumptions often come with weaker certification.

A particularly relevant differentiator is constraint handling.
Among the surveyed methods, \paperNameMPC{} is the only one that supports state and input constraints in the synthesis and execution loop.
This is a major advantage in safety-critical applications, where respecting bounds can be as important as convergence.
The absence of explicit constraint handling does not make the other methods unusable, but they may require wrappers such as safety filters or conservative tuning.

\begin{figure*}[t]
    \tikzexternaldisable
    \centering
    \includegraphics{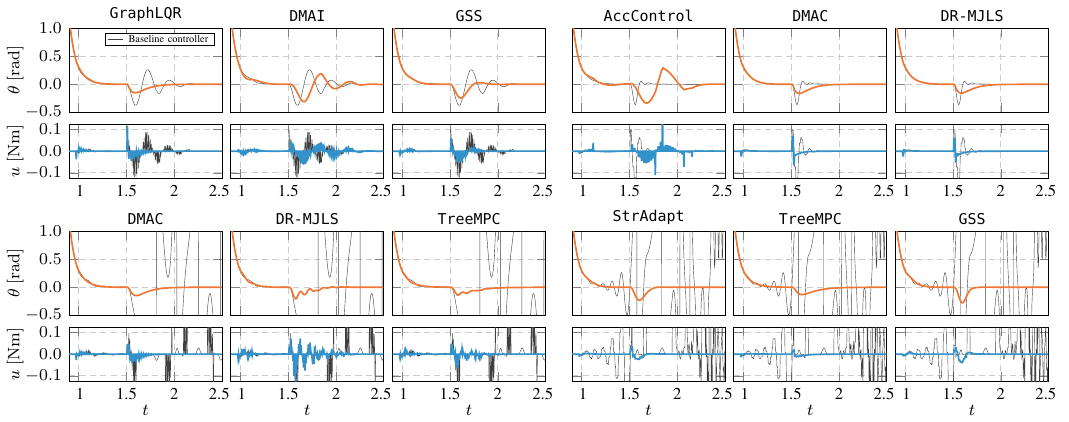}
    \vspace{-0.5cm}
    \caption{Furuta pendulum: Trajectories of pendulum angle $\theta(t)$ and system input $u(t)$ with a common sequence $(\mu[k])$, using \Kill{}/\Zero{} (top left), \Kill{}/\Hold{} (top right), \Skip{}/\Zero{} (bottom left) or \Skip{}/\Hold{} (bottom right).}
    \label{fig:furuta}
    \vspace{-4mm}
\end{figure*}

\subsection{Summary}
\label{sec:qualitative-summary}

No method dominates across all dimensions.
Methods with broad execution-semantics support and strong guarantees typically require richer timing models and heavier offline design, whereas methods that are easier to retrofit into existing systems usually provide weaker formal certification or less expressive optimization objectives.
This supports the main thesis of the paper: deadline-miss-aware controller design is best understood as a design space with trade-offs, rather than a search for a single best solution.

\section{Experimental results}
\label{sec:experiments}

To complement the qualitative comparison, we propose a simulation framework for quantitative scenario-specific comparisons of the controllers (Sec.~\ref{sec:experiments:framework}).
To illustrate its capabilities, the framework is tested over a set of experiments on two target benchmarks, a Furuta pendulum (Sec.~\ref{sec:experiments:furuta}) and an electric motor (Sec.~\ref{sec:experiments:pmsm}),  showing examples of the effects of different design constraints and deadline-miss models.
We refrain from comparing the surveyed methods in absolute terms, based only on the shown experiments: The results of the following comparisons are to be read as specific for the selected use-cases, deadline-miss scenarios, and control tuning.

\subsection{Simulation framework}
\label{sec:experiments:framework}

We implemented the comparison framework and all controllers in MATLAB\textsuperscript{\textregistered}.
The code (publicly available\footnote{\href{https://github.com/martinamaggio/missaware-controller-synthesis}{https://github.com/martinamaggio/missaware-controller-synthesis}.}) has been designed with a modular organization to facilitate adding and evaluating controllers in a common simulation setup.

The project comprises four subfolders.
The \texttt{model} folder contains the plant and timing-constraint models and the simulation routines.
Users can choose one of the provided plant examples, or add their own models and parameters.
The \texttt{metrics} folder contains comparison indicators, including standard tracking and regulation measures and task-specific criteria such as maintaining the pendulum near the upright equilibrium.
The \texttt{controllers} folder contains the implementations discussed in Sec.~\ref{sec:comparison} (with optimization problems solved using the interface YALMIP \cite{lofberg2004yalmip}), and the \texttt{experimental} folder contains the scripts and configuration files for the evaluation campaigns.
Together, these modules allow closed-loop simulation under selected plants, controllers, deadline hit/miss patterns, and overrun and actuation strategies, and are used to produce the results reported next.

\subsection{Furuta pendulum}
\label{sec:experiments:furuta}

\begin{figure*}[ht]
    \tikzexternaldisable
    \centering
    \includegraphics{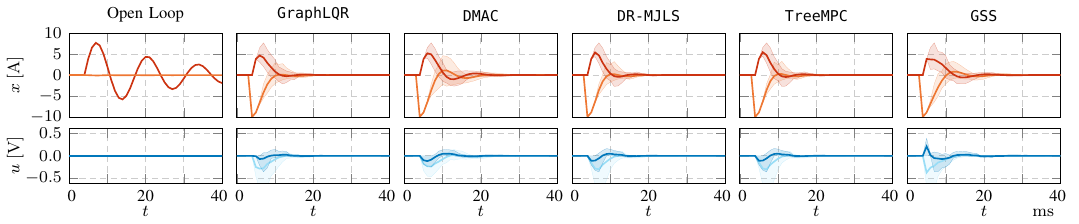}
    \vspace{-0.5cm}
    \caption{Electric motor: State trajectories $i_\mathrm{d}(t)$ ({\protect\tikz[baseline=-0.5ex] \protect\draw[colfirststate,thick] (0,0) -- (0.35,0);}), $i_\mathrm{q}(t)$ ({\protect\tikz[baseline=-0.5ex] \protect\draw[colsecondstate,thick] (0,0) -- (0.35,0);})
    and inputs $u_\mathrm{d}(t)$ ({\protect\tikz[baseline=-0.5ex] \protect\draw[colfirstinput,thick] (0,0) -- (0.35,0);}), $u_\mathrm{q}(t)$ ({\protect\tikz[baseline=-0.5ex] \protect\draw[colsecondinput,thick] (0,0) -- (0.35,0);}) with offset disturbance, using \Kill{} and \Zero{}. The 0.1--0.9 quantiles are shown with opacity. The trajectories of $\omega_\mathrm{el}(t)$ are omitted.}
    \label{fig:kill_zero_motor}
    \vspace{-4mm}
\end{figure*}

\begin{figure}
    \tikzexternaldisable
    \centering
    \includegraphics{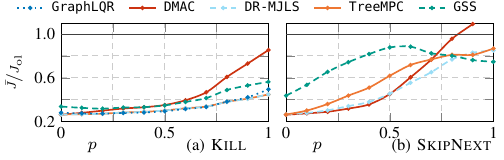}
    \vspace{-0.1cm}
    \caption{Electric motor: Cost ratio $\nicefrac{\Bar{J}}{J_\mathrm{ol}}$ against different probabilities $p$ of deadline misses.}
    \label{fig:varying_p_miss}
    \vspace{-4mm}
\end{figure}

First, we present an overview of the framework's capabilities on a comparable simulation setting for all selected methods.
We use a Furuta pendulum as an academic framework, simulated using a state-of-the-art nonlinear model from~\cite{Gafvert1998modelling}\footnote{
    The model parameters are: pendulum arm length $0.075\,$m, base radius $0.043\,$m, gravity $9.81\,\text{m}/\text{s}^2$, pendulum mass $0.0054\,$kg and moment of inertia $125.1\cdot 10^{-6}\,\text{kg}\,\text{m}^2$. These parameters resemble the physical pendulum used in one of the most recent synthesis papers~\cite{seidel2026framework}.%
    } 
of the form in Eq.~\eqref{eq:plant:nonlinear}.
For controller synthesis, the model is linearized around the upright equilibrium  in the form of Eq.~\eqref{eq:sys-ct}, then discretized with period $T=5\,\mathrm{ms}$ in the form of Eq.~\eqref{eq:sys-dt}.
The state $x=[\theta,\dot{\theta},\dot{\phi}]$ consists of pendulum angle $\theta$, pendulum angular velocity $\dot{\theta}$, and base angular velocity $\dot{\phi}$, while the input $u$ is the torque applied at the base.

We present one example simulation of all eight controllers (plus a baseline controller with no deadline-miss awareness, from \cite{Josephrexon2022}), with different overrun and actuator strategy combinations, subject to an offset disturbance.
With the goal of fulfilling the assumptions on the deadline-miss model for all methods, a sequence of outcomes is generated that complies with the periodic hit pattern required by \paperNameGaid{} and with the corresponding \anyHit{\nhit}{\window} constraint or the over-approximating \rowMiss{\nmiss} constraint, where $\nhit=2$, $\nmiss=3$, and $\window=5$.
The Markov graph of the corresponding patterns is used for the synthesis of \paperNameDMAC{}, substituting the scenario theory step. 
The outcomes of periods that are not required to be hits by the periodic hit model are generated probabilistically and independently, with a deadline-miss probability of $p=0.9$.

We show all controllers at least once\footnote{More comprehensive figures including a second disturbance scenario with brown process noise on $\theta(t)$ are available in the linked GitHub repository.} in Fig.~\ref{fig:furuta}, grouped according to their respective strategy combination.
The figure shows base torque $u(t)$ and pendulum angle $\theta(t)$
in the region around the upright equilibrium, where the deadline-miss-aware balancing controllers are active, after a swing-up event.
The results are shown under a disturbance scenario, where an offset of $\Delta = 4\, \nicefrac{\mathrm{rad}}{\mathrm{s}}$ is added to $\dot{\phi}(t)$ at $t=1.5\,\mathrm{s}$ for two time steps.

The controllers with deadline-miss awareness are all stabilizing under this specific sequence of outcomes, while the system becomes unstable with the baseline controller under the \Skip{} strategy.
However, in our tests, the controllers show different robustness.
When selecting different \gls{wh} constraints parameters, the methods that explicitly take into account such knowledge in the design phase to provide stability guarantees perform more consistently.
On the other hand, the performance of both \paperNameStreamAdapt{} and \paperNameVreman{}, where stability is considered only a-posteriori, was strongly linked to the initial tuning of the nominal adapted controller.
Additionally, shifting the given outcome sequence by just one or two steps can deteriorate the performance of \paperNameGaid{}.
As discussed in Section~\ref{sec:qualitativecomparison}, selecting the optimal controller is highly application-dependent and this limits the possibility of performing a universally fair quantitative comparison.

\begin{figure*}
    \tikzexternaldisable
    \centering
    \includegraphics{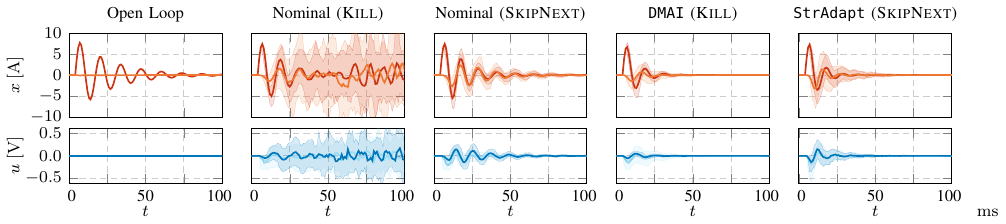}
    \vspace{-0.2cm}
    \caption{Electric motor: Trajectories and 0.1--0.9 quantiles of states $i_\mathrm{d}(t)$ ({\protect\tikz[baseline=-0.5ex] \protect\draw[colfirststate,thick] (0,0) -- (0.35,0);}), $i_\mathrm{q}(t)$ ({\protect\tikz[baseline=-0.5ex] \protect\draw[colsecondstate,thick] (0,0) -- (0.35,0);})
    and inputs $u_\mathrm{d}(t)$ ({\protect\tikz[baseline=-0.5ex] \protect\draw[colfirstinput,thick] (0,0) -- (0.35,0);}), $u_\mathrm{q}(t)$ ({\protect\tikz[baseline=-0.5ex] \protect\draw[colsecondinput,thick] (0,0) -- (0.35,0);}) controlled by output-feedback at an upper bound of the job response time $\bar{t} = 2.5\,T$. The trajectories of $\omega_\mathrm{el}(t)$ are omitted.}
    \label{fig:output_feedback_trajectories}
    \vspace{-4mm}
\end{figure*}

\begin{figure}
    \tikzexternaldisable
    \centering
    \includegraphics{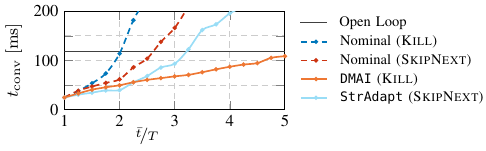}
    \vspace{-0.25cm}
    \caption{Electric motor: Convergence times $t_\mathrm{conv}$ against different upper bounds of the job response times $\bar{t}$.}
    \label{fig:output_feedback_metrics}
    \vspace{-4mm}
\end{figure}

\subsection{Electric motor}
\label{sec:experiments:pmsm}

In this section, we consider a \gls{pmsm}, a common type of electric motor, as industry-relevant use-case.
The \gls{pmsm} is modeled\footnote{The model parameters are: winding resistance $0.01\,\Omega$, inductance in \textit{d} and \textit{q} directions $1\cdot10^{-4}\,\mathrm{H}$ and $1.2\cdot10^{-4}\,\mathrm{H}$, respectively, permanent magnet flux linkage $0.05\,\mathrm{Wb}$, number of pole pairs $6$, and inertia $0.005\,\mathrm{kg\,m^2}$.} using the differential equations from \cite{kiselev2018pmsm} for a system state $x=[i_\mathrm{d},i_\mathrm{q},\omega_\mathrm{el}]^\top$, including currents in rotating \textit{dq}-coordinates and electrical rotational speed, and input $u=[u_\mathrm{d},u_\mathrm{q}]^\top$ comprising the applied voltages in \textit{dq}-coordinates.
The controllers are implemented with $T=1\,\mathrm{ms}$.
With our framework, multiple case studies can be defined and the available controllers compared:
We consider two case studies, highlighting how different factors may affect which controller design method is the most suitable.

\vspace{2mm}
\noindent\textit{Influence of the deadline-miss probability.}
In this case study, we investigate the difference between stochastic and deterministic approaches.
We focus on methods that both explicitly take into account a deadline-miss model at design time, and support a \rowMiss{4} constraint with \Zero{} actuation strategy (\Hold{} is harder to implement for \glspl{pmsm} due to the rotating reference frame). 
Whenever the \gls{wh} constraint allows for a deadline miss, its probability to happen is $p$.
We analyze how different values of $p$ influence the control performance under different controller design methods, and if additional knowledge on those probabilities improves it.
We run Monte Carlo simulations of $200$ sample sequences for each value of $p$, extracting the quadratic costs $J$ of each closed-loop control system, with identity matrices for $Q_\mathrm{x}, Q_\mathrm{u}$ (c.f., Box~\ref{box:cost-functions}).
Since the \gls{pmsm} is open-loop stable, we compare the closed-loop performances against the open-loop quadratic cost $J_\mathrm{ol}$.
At $t = 3\,\mathrm{ms}$, an offset is added to $\omega_\mathrm{el}$, representing a sudden change in the load torque.

Fig.~\ref{fig:kill_zero_motor} shows the system's state and input trajectories under the selected controllers for \Kill{} and \Zero{}, with $p = 0.5$: 
In this case, all controllers outperform the open-loop system, despite deadline misses.
The averaged cost $\bar{J}$ relative to $J_\mathrm{ol}$ plotted over $p$ is depicted in Fig.~\ref{fig:varying_p_miss} for both \Kill{} and \Skip{} overrun strategies.
As expected, the costs increase with larger $p$, as more deadline misses typically lead to worse performance.
Nonetheless, the controllers still outperform $J_\mathrm{ol}$  for nearly all combinations of designs and probabilities $p$.
Furthermore, stochastic design methods like \paperNameMJLS{} can achieve lower costs compared to deterministic ones for some values of $p$.
This is due to the inclusion of the knowledge on deadline-miss probabilities in their design, while deterministic designs are optimized for worst case scenarios---closer to the conditions obtained with large $p$.
All selected controller designs use quadratic costs matching $J$, except for \paperNameMS{}, which optimizes $\ell_2$-performance.
Furthermore, \paperNameMPC{} has a significantly higher computation time (of one to two orders of magnitude) in this setting, however it guarantees constraint satisfaction.
However, all controllers can be tuned to some extent and can behave differently for other cost functions.

\vspace{2mm}
\noindent\textit{Output feedback.}
This case study assumes that only partial state measurement  ($i_\mathrm{d}(t)$ and $\omega_\mathrm{el}(t)$) is possible.
Therefore, only \paperNameStreamAdapt{} and \paperNameVreman{} are applicable.
Since both works are based on adapting a given nominal controller, we also include it in the comparison.

\paperNameVreman{} and \paperNameStreamAdapt{} both allow for \rowMiss{m} and \Hold{}, but require two different overrun strategies, namely \Kill{} and \Skip{}, respectively.
In order to provide a comparable real-time setting, we first generate a sequence of sampled response times and derive the corresponding sequence of outcomes for both \Kill{} and \Skip{}, while ensuring a bounded value for $m$.
Each periodic job's response time is sampled from an uniform distribution with lower bound $0.1\,T$ and parametric upper bound $\Bar{t} \geq T$.
We run a Monte Carlo simulation with $200$ sample sequences each for increasing values of $\Bar{t}$, adding the same disturbance to $\omega_\mathrm{el}$ as above.

The example demonstrates that it is still possible to achieve control goals despite the absence of full state measurements and to improve performance by adapting a control policy to account for deadline misses.
Fig.~\ref{fig:output_feedback_trajectories} presents the mean and quantiles of the trajectories with $\Bar{t} = 2.5\,T$, for the selected controllers and overrun strategies, showing faster convergence of the deadline-miss-aware designs.
Fig.~\ref{fig:output_feedback_metrics} shows the mean convergence time for each combination, as the duration between the application of the offset disturbance and the point of time where $|x_i(t)| < 0.05 \, \forall i$.
Since all controllers are based on the same nominal one, they yield identical results if no deadlines are missed.
However, for an increasing value of $\bar{t}$, the plots show a stronger deterioration of the nominal controllers' performance, particularly under the \Kill{} strategy.
Indeed, when sampling response times from a distribution, \Skip{} nonetheless ensures that the delayed job will eventually complete, while under \Kill{} there might be a larger number of consecutive intervals without updates.
Remarkably, Fig. \ref{fig:output_feedback_metrics} still indicates the best performance and an improvement over the open loop convergence at all points using \paperNameVreman{} and \Kill{}, with \paperNameStreamAdapt{} (originally targeted at systems with sensor periods smaller than their control period) generally above, but below the nominal controllers.

\section{Related work}
\label{sec:related}

Controller co-design under real-time constraints has been widely studied~\cite{arzen2000introduction, zhu2018codesign}.
Existing works address robustness to jitter~\cite{cervin2004jitter, buttazzo2007comparative, aminifar2013control}, exploit response-time analysis for performance improvement~\cite{xu2015exploiting}, and incorporate timing into co-simulation tools such as Jitterbug, TrueTime~\cite{cervin2003does}, and Jittertime~\cite{cervin2019using}.
Other lines of work consider the scheduling of overloaded control systems, including feedback scheduling~\cite{cervin2002feedback}, elastic tasks~\cite{buttazzo2007quality}, and job skipping~\cite{yoshimoto2011optimal, wang2021cross}.
However, these approaches mostly fall outside the scope of this paper, since they focus on schedulability while assuming pre-designed controllers.

A broad literature also studies control under network delays and losses, see~\cite{zhang2016ncssurvey}.
The connection between deadline misses and packet dropouts is well known: under the \Kill{} strategy, a dropped packet is equivalent to a terminated job, since the receiver has no updated command.
By contrast, overrun strategies such as \Skip{} and \Queue{} are rarely considered.

To capture such effects more expressively, several timing abstractions have been proposed, including soft deadlines~\cite{fontanelli2013soft}, the ($h,w$)-firm model~\cite{ramanathan2002overload}, and the \gls{wh} model~\cite{bernat2001weakly}.
These abstractions have been used in co-design for overloaded real-time control systems and packet dropouts~\cite{gaid2008dropin, jia2005analysis, flavia2006impact, vanhorssen2016performance}.

More recently, the \gls{wh} model has attracted growing attention due to its favorable trade-off between expressivity and modeling simplicity~\cite{gujarati2019iteration, zhu2020know}.
In real-time control, deadline-miss-aware formulations were first investigated in~\cite{frehse2014formal, geelen2015impact}.
\Gls{wh} constraints have been first applied to packet-drop modeling in \gls{ncs} in~\cite{blind2015whnetwork},
later extended through graph-based representations in~\cite{linsenmayer2017stabilization, linsenmayer2021stabilization, seidel2024l2performance}.
Controllers subject to deadline misses have been analyzed for stability requirements~\cite{goswami2014relaxing, liang2019security, maggio2020control, vreman2022extended}, performance criteria~\cite{pazzaglia2018beyond, pazzaglia2019dmac, liang2020leveraging, vreman2021stability}, safety~\cite{huang2019formal,huang2020saw}, security~\cite{liang2019security} and fault tolerance~\cite{liang2020leveraging}.
While a broad overview is available in~\cite{salamun2023weakly}, our paper focuses on a detailed analysis and comparison of selected deadline-miss-aware control design approaches.

\section{Conclusion}
\label{sec:conclusions}

This paper provides a qualitative comparison of representative deadline-miss-aware controller design methods, together with a modular framework for their systematic evaluation.
We illustrate how the considered methods differ in assumptions, runtime semantics, guarantees, and implementation effort.
We also provide multiple examples of usage of the simulation framework under different deadline-miss conditions and system assumptions, underlining the qualitative results.
The main takeaway is that deadline-miss-aware controller design is best understood as a trade-off space rather than a search for a universally best solution.
We believe the framework and analysis presented in this paper will facilitate more reproducible comparisons and help position future contributions in this area.%

\section*{Acknowledgments}
We acknowledge the use of generative AI tools (Gemini 3.1 Pro) solely to assist with formatting of TikZ figures.

\bibliographystyle{IEEEtran}
\bibliography{biblio}

\end{document}